\documentclass[%
 preprint,
superscriptaddress,
 amsmath,amssymb,
 aps,
]{revtex4-1}

\usepackage{graphicx}% Include figure files
\usepackage{dcolumn}% Align table columns on decimal point
\usepackage{bm}% bold math
\begin{document}

\preprint{APS/123-QED}

\title{Two-Photon Rabi Splitting in a Coupled System of a Nanocavity and Exciton Complexes}
%Correlated Multi-Transition Cavity Quantum Electrodynamics System With Large Coupling Strength}

\author{Chenjiang Qian}
\author{Shiyao Wu}
\author{Feilong Song}
\author{Kai Peng}
\author{Xin Xie}
\author{Jingnan Yang}
\author{Shan Xiao}
\affiliation{Institute of Physics, Chinese Academy of Science, Beijing 100190, China}
\affiliation{School of Physical Sciences, University of Chinese Academy of Sciences, Beijing 100049, China}
\author{Matthew J. Steer}
\author{Iain G. Thayne}
\affiliation{School of Engineering, University of Glasgow, Glasgow G12 8LT, U.K.}
\author{Chengchun Tang}
\author{Zhanchun Zuo}
\affiliation{Institute of Physics, Chinese Academy of Science, Beijing 100190, China}
\author{Kuijuan Jin}
\author{Changzhi Gu}
\affiliation{Institute of Physics, Chinese Academy of Science, Beijing 100190, China}
\affiliation{School of Physical Sciences, University of Chinese Academy of Sciences, Beijing 100049, China}

\author{Xiulai Xu}
% \altaffiliation[Also at ]{Institute of Physics, Chinese Academy of Sciences, Beijing.}%Lines break automatically or can be forced with \\
\email{xlxu@iphy.ac.cn}
\affiliation{Institute of Physics, Chinese Academy of Science, Beijing 100190, China}
\affiliation{School of Physical Sciences, University of Chinese Academy of Sciences, Beijing 100049, China}
\affiliation{CAS Center for Excellence in Topological Quantum Computation, University of Chinese Academy of Sciences, Beijing 100190, China}


%\affiliation{
%\affiliation{
%\affiliation{%


\date{\today}% It is always \today, today,

\begin{abstract}

Two-photon Rabi splitting in a cavity-dot system provides a basis for multi-qubit coherent control
in quantum photonic network. Here we report on two-photon Rabi splitting in a strongly coupled cavity-dot system. The quantum dot was grown intentionally large in size for large oscillation strength and small biexciton binding energy. Both exciton and biexciton transitions couple to a high quality factor photonic crystal cavity with large coupling strengths over 130 $\mu$eV. Furthermore, the small binding energy enables the cavity to simultaneously couple with two exciton states. Thereby two-photon Rabi splitting between biexciton and cavity is achieved, which can be well reproduced by theoretical calculations with quantum master equations.

\begin{description}
\item[PACS numbers]
42.50.Pq,78.67.Pt,78.67.Hc
\end{description}
\end{abstract}

\maketitle

%\tableofcontents
%\section{\label{sec1}Introduction:\protect\\}

%and has many significant applications such as entangled photon generation \cite} and two-photon laser \cite{}.
%\cite{PhysRevLett.100.240404,Müller2014,PhysRevLett.59.1899,PhysRevLett.93.187403}

Two-photon process in quantum electrodynamics is important for investigating light-matter interaction. Similar to single-photon process, two-photon Rabi oscillation occurs when two-photon exchange rate between an emitter and electromagnetic field exceeds their decay rates, providing a basis for multi-photon coherent control \cite{LIANG2006171,0953-4075-41-8-085402,RevModPhys.82.2313,Fushitani2015,PhysRevLett.118.233601}. A single quantum dot (QD), containing exciton (X) and biexciton (XX) states, could serve as a two-photon emitter \cite{PhysRevB.73.125304} with long coherence time \cite{wang2007self}. The coupled cavity-dot system can be used as a basic building block of quantum photonic network \cite{nakamura2004ultra,angelakis2007proposal,faraon2008dipole,bose2012low,PhysRevLett.112.213602}. However, two-photon Rabi splitting in a cavity-dot system has not yet been experimentally demonstrated, restricting its applications in multi-photon operation. This is due to that the biexciton binding energy of QDs is too large in general and coupling between two-photon transition and cavity mode is not strong enough \cite{PhysRevB.81.035302,PhysRevB.95.245306}. To achieve strong-coupling regime in a cavity-dot system, a promising way is to utilize photonic crystal (PC) cavity, due to high quality factor (Q) and small mode volume (V) \cite{akahane2003high,vahala2003optical,yoshie2004vacuum,hennessy2006quantum}.

In the past decade, PC based cavity-dot system has been continuously optimized for larger coupling strength and more nonlinearity features \cite{strauf2006self,brossard2010strongly,PhysRevLett.108.093604,kim2013quantum,sun2016quantum}. Nonetheless, up to now studies have been mainly focused on a single transition and a single cavity mode. Recently, a few investigations were reported on coupled systems between cavity and two transitions from one single QD \cite{winger2008quantum,ota2011spontaneous} or two different QDs\cite{PhysRevB.82.075305,kim2011strong}. Ota et al \cite{ota2011spontaneous} demonstrated two-photon emission enhancement, based on two single-photon strong couplings with coupling strength of 51 (43) $\mu$eV between the cavity and exciton (biexciton) state from a single QD.

%QDs contain multiple transitions and could implement both exciton \cite{stievater2001rabi} and spin \cite{petta2005coherent,press2008complete,Mar2014} qubits.

In this letter, we report on two-photon Rabi splitting in a strongly coupled cavity-dot system consisting of a nanocavity and two exciton states (X and XX) from a single QD. The obtained single-photon coupling strengths are about 130 $\mu$eV, twice over the previous value \cite{ota2011spontaneous}, which is due to the large oscillation strength and the large wave function overlapping with cavity mode, resulting from relatively large size of QDs \cite{doi:10.1063/1.4965845,PhysRevB.93.155301}. Meanwhile quantum confinement is weak in large QDs, leading to a small binding energy \cite{PhysRevB.46.15578,PhysRevLett.64.1805}. These made the cavity simultaneously couple to two single-photon transitions, resulting in two-photon Rabi splitting between biexciton and cavity, which is well explained by theoretical simulations.

\begin{figure}
\includegraphics[scale=0.43]{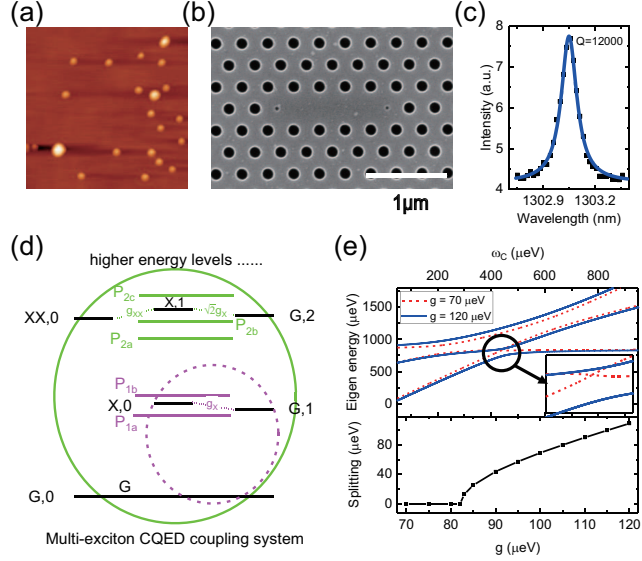}
\caption{\label{f1} (a) Atomic force microscope image of QDs in 1$\mu m^2$. Diameter of the biggest QD is around 50 nm. (b) Scanning electron microscope image of L3 PC cavity. Two red circles schematize the positions of two edge holes in an unmodified L3 cavity. The two edge holes were optimized by shift of 0.15 a and shrink of 0.13 a. (c) One cavity mode measured at room temperature with Q=12000. (d) Energy level structure of the cavity-dot system with large coupling strength. (e) (upper panel) Calculated eigen energies of three polaritons from coupling between $|XX,0\rangle$, $|X,1\rangle$ and $|G,2\rangle$, at g=70$\ \mu eV$ (red dashed line) and at g=120$\ \mu eV$ (blue solid line). (inset) Two-photon Rabi splitting occurs when g=120$\ \mu eV$. (bottom panel) Two-photon Rabi splitting energy as a function of g.}
\end{figure}

%\section{\label{sec2}Devices and Methods:\protect\\}
L3 PC cavities with various parameters were fabricated on a 170 nm thick GaAs slab. InAs QDs with a density of ~10$^9$ cm$^{-2}$ were grown in the middle. The QDs were grown at a quite low growth rate to allow better control of the thickness for low density and achieve large dot in size (Fig.~\ref{f1}(a)) for small binding energy and large oscillation strength. The temperature-dependent PL measurement was performed with a conventional confocal micro-PL setup. The overall cavity Q is around 10000, which is high enough to achieve strong coupling in cavity-dot system\cite{yoshie2004vacuum,hennessy2006quantum}. Fig.~\ref{f1}(b) shows a typical cavity and Fig. 1(c) shows a typical cavity mode with Q=12000 fitted with Lorentzian shape, which could be higher after deconvolution \cite{jimenez1994contribution}. The details of fabrication and measurement are shown in Supplementary Materials.

The energy level structure of coupled cavity and biexciton system is shown in Fig.~\ref{f1}(d). Each energy level contains the QD state and photon number in cavity. Ground state $|G,0\rangle$ is labelled by G. Single exciton state $|X,0\rangle$ couples to single photon state $|G,1\rangle$ with $g_{X}$, forming two polaritons labelled by $P_{1a}$ and $P_{1b}$. Single exciton with one photon state $|X,1\rangle$ couples to both biexciton state $|XX,0\rangle$ with $g_{XX}$ and two-photon state $|G,2\rangle$ with ${\sqrt{2}g_{X}}$, forming three polaritons labelled by $P_{2a}$,  $P_{2b}$ and $P_{2c}$. Similar to a three-level system in atoms \cite{0953-4075-41-8-085402}, two-photon Rabi oscillation could be observed between $|XX,0\rangle$ and $|G,2\rangle$ when they are close to resonance, along with large coupling strengths ($g_{X}$, $g_{XX}$) and small biexciton binding energy. To understand this model, first we introduce a single-photon-exciton coupling system in the limit of weak excitation \cite{PhysRevA.40.5516}, with energy levels highlighted in the purple dashed circle in Fig.~\ref{f1}(d). Coupling between $|X,0\rangle$ and $|G,1\rangle$ can be described by Hamiltonian matrix
\begin{eqnarray}
\left(
\begin{array}{cc}
\omega_{X}+\frac{i\gamma_X}{2}&g_X\\
\\
g_X&\omega_{C}+\frac{i\kappa}{2}
\end{array}\right)\;,
\end{eqnarray}
\\
where $\omega_{X}$ and $\omega_{C}$ are the eigenfrequencies of the exciton state and the cavity mode, $\gamma_X$ and $\kappa$ correspond to the cavity loss and decay rate of X transition, respectively. Two eigenvalues including energy and decay rate of $P_{1a}$ and $P_{1b}$ are $(\omega_{X}+\omega_{C})/2+i(\gamma_X+\kappa)/4\pm\sqrt{g^{2}+(1/4)[\omega_{X}-\omega_{C}+i(\gamma_X-\kappa)/2]^{2}}$ \cite{andreani1999strong}. Strong coupling occurs when $g>(\kappa-\gamma_X)/4$. Then we move to the biexciton system in the limit of weak excitation, as highlighted in green solid circle. Coupling between $|XX,0\rangle$, $|X,1\rangle$ and $|G,2\rangle$ can be described by Hamiltonian matrix
\begin{eqnarray}
\left(
\begin{array}{ccc}
\omega_{XX}+\omega_{X}+i\frac{\gamma_{XX}}{2}&g_{XX}&0\\
g_{XX}&\omega_{X}+\omega_{C}+i\frac{\gamma_X+\kappa}{2}&\sqrt{2}g_X\\
0&\sqrt{2}g_X&2\omega_{C}+i\kappa
\end{array}\right)\;.\ \ \ \
\end{eqnarray}
Here, $\omega_{X}+\omega_{XX}$ and $\gamma_{XX}$ represent the eigenfrequency and the decay rate of biexciton state respectively. Analytical eigenvalues of $3 \times 3$ Matrix are very complex. Instead we chose a set of parameters $\omega_{X}=600\ \mu eV$, $\omega_{XX}=250\ \mu eV$, $\gamma_{XX}=\gamma_{X}=5\ \mu eV$, $\kappa=90\ \mu eV$ to simulate numerical eigenvalues as $\omega_{C}$ change from 0 to 1000 $\mu eV$ with various $g_X=g_{XX}=g$. Upper panel in Fig.~\ref{f1}(e) shows simulated energies of $P_{2a}$,  $P_{2b}$ and $P_{2c}$ at coupling strength $g=70\ \mu eV$ (red dashed line) and $g=120\ \mu eV$ (blue solid line). Single-photon Rabi splitting occurs between cavity and two single-photon transitions under both conditions. While two-photon Rabi splitting occurs only at $g=120\ \mu eV$, when $|XX,0\rangle$ and $|G,2\rangle$ are close to the resonance (as magnified in the inset in Fig.~\ref{f1}(e)). Bottom panel in Fig.~\ref{f1}(e) shows two-photon Rabi splitting energy at different g values, indicating a threshold value of $82\ \mu eV$. Specific PL spectra simulated by solving master equation using Quantum Optics Toolbox \cite{tan2002quantum} with different coupling strengths are shown in Supplemental Materials.

\begin{figure}[b]
\includegraphics[scale=0.43]{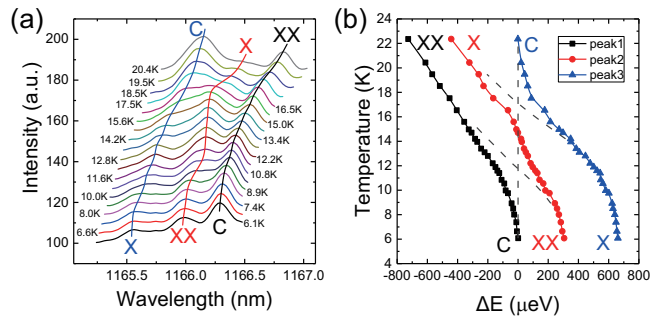}
\caption{\label{f2} (a) PL spectra of X and XX transitions and a cavity mode collected from 6 K to 20 K with the excitation power of 500 nW. Peak1, peak2 and peak3 are colored coded in black, red and blue, respectively. (b) Detuning between three peaks and bare cavity mode as a function of temperature. The detunings between uncoupled QDs and cavity are indicated by brown dashed lines.}
\end{figure}

Temperature-dependent PL spectra (Fig.~\ref{f2}(a)) were collected with an excitation power of 500 nW, under which both excitonic transitions and cavity mode could be observed. As temperature increases, the shift of QD emission energy could be mainly ascribed to band-gap shrinkage of the InAs QDs, following the empirical Varshni relation \cite{varshni1967temperature,lee1997temperature}. However, the cavity mode energy shift is affected by two mechanisms. One is the increase of bulk refractive index, leading to a red shift \cite{yoshie2004vacuum}. The other one is the evaporating of condensed residual gas on the sample surface, resulting in a blue shift \cite{brossard2010strongly,mosor2005scanning}. Fig.~\ref{f2}(a) consists of three distinguishable peaks originating from transitions between the states formed by coherent coupling between QD excitons and cavity mode. Peak1 is identified as cavity mode while peak2 and peak3 are two QD transitions at 6 K. Meanwhile peak3 is denoted as cavity mode while peak1 and peak2 as two QD transitions at 20 K. Fig.~\ref{f2}(b) shows the detunings (solid lines) between three peaks and bare cavity mode as a function of temperature and comparing with uncoupled QD transitions and a bare cavity (dashed lines). PL spetrum fitting and temperature-dependent characterization of bare cavity and QDs are shown in the Supplementary Materials. Clearly, two anti-crossing behaviors with vacuum Rabi splitting of 246 $\mu$eV at 11 K and 242 $\mu$eV at 17 K were observed, indicating large strong coupling between the cavity and two transitions.

\begin{figure}
\includegraphics[scale=0.43]{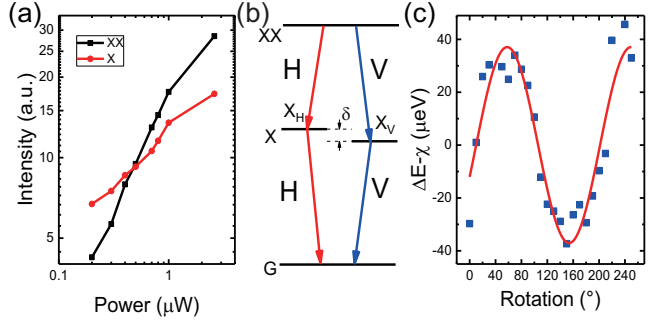}
\caption{\label{f3} (a) Peak intensity as a function of excitation power. Peak2 is weaker than peak3 at low excitation power but grows faster with increasing pumping power. (b) Four-level energy structure of QD. X state consists of X$_H$ and X$_V$ with a fine structure splitting energy $\delta$. (c) Energy difference at different linear polarization angles. Solid red line shows a Sinusoidal fitting with a $\pi$ period and the fitted fine structure splitting energy is around 37 $\mu$eV .}
\end{figure}

\begin{figure}[b]
\includegraphics[scale=0.43]{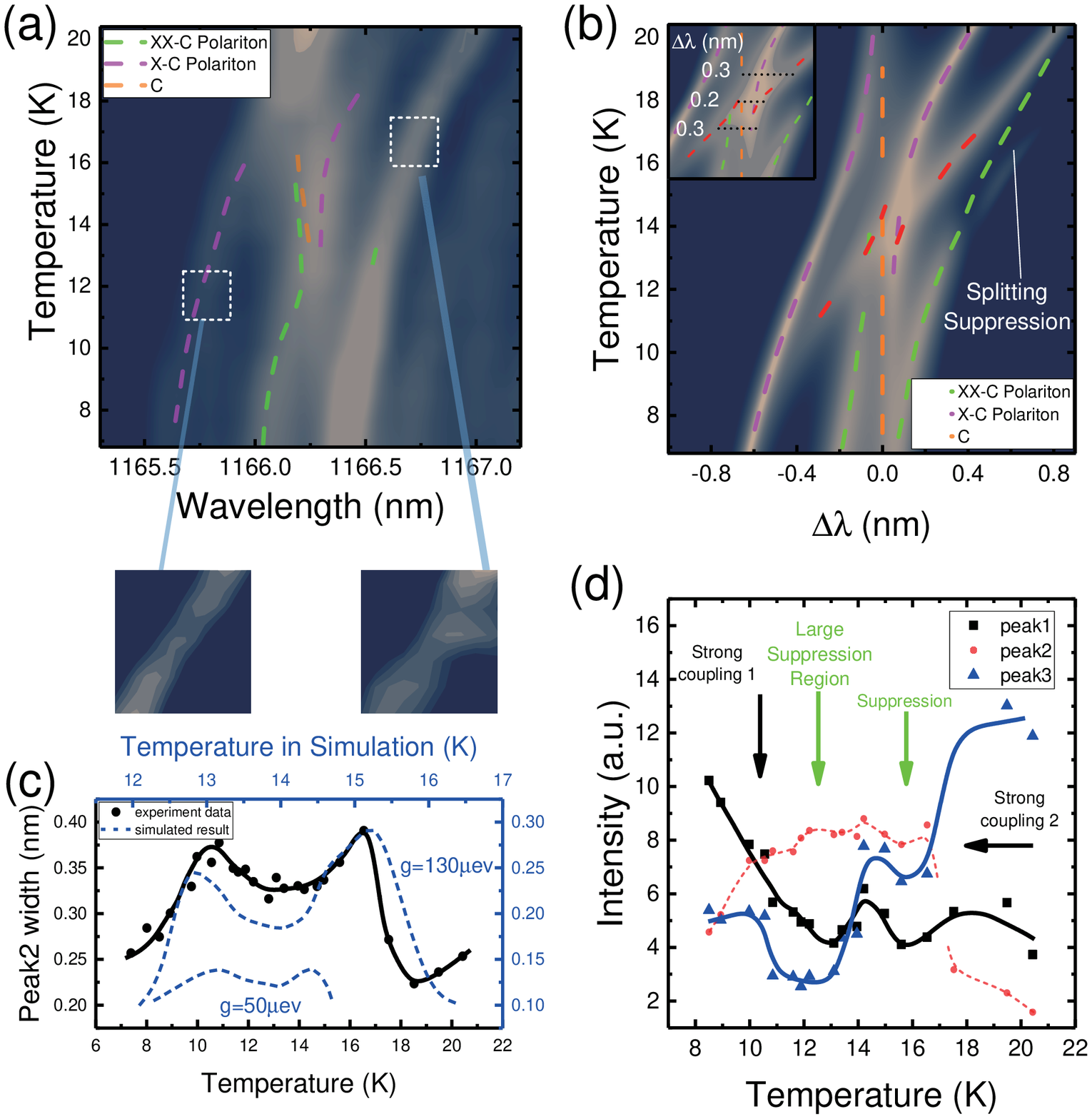}
\caption{\label{f4} (a) Contour plot of PL spectra with two suppression regions, as shown in the insets. (b) Simulated PL map with g=130 $\mu$eV. Green dashed line represents the XX-C polariton, purple dashed line shows the X-C polariton and orange dashed line is bare cavity. Red dashed line is a polariton-polariton transition contains two-photon Rabi splitting. (inset) Linewidth variation during two-photon Rabi splitting. (c) Linewidth variation of peak2 at different temperatures in experiment (solid line) and calculation (dashed lines). The difference of $\sim 0.1~nm$ between experimental data and calculation with g=130 $\mu$eV results from the broadening of the spectrometer. (d) Intensity variation of experimental data at different temperatures. Large suppression region results from two-photon Rabi splitting.}
\end{figure}

Strongly coupled to a cavity, the two QD peaks might originate from two QDs \cite{PhysRevB.82.075305,kim2011strong} or different transitions of one single QD \cite{winger2008quantum,ota2011spontaneous}. To identify the two QD peaks, we first measured the excitation power-dependent PL of peak2 and peak3 at 6 K (Fig.~\ref{f3}(a)). The intensity of peak2 is lower than peak3 at low excitation power, but grows faster with increasing excitation power. The slope of the two lines plotted in logarithm is $k_{XX}=0.90$ and $k_{X}=0.43$. $k_{XX}/k_{X}=2$ manifests the characteristic of XX and X transitions of a single QD. $k_{XX}$($k_{X}$) is smaller than the value of 2(1) at extreme low excitation power \cite{PhysRevLett.73.1138}, which is might due to that emission of QD is close to saturation in our work. The energy difference $\Delta E$ between two peaks comes from the binding energy $\chi$=~350 $\mu$eV, quite small comparing with typical InAs QDs \cite{PhysRevB.81.035302,wang2007self,PhysRevB.95.245306}. Then a fine structure splitting measurement was applied to confirm our assumption. The fine structure splitting comes from asymmetry in pyramidal structure of self-assembled QDs \cite{gammon1996fine}, as the energy-level diagram shown in Fig.~\ref{f3}(b). The polarization-resolved PL measurement should show an oscillation of $\Delta E$ with an amplitude of $\delta$ between the two orthogonal linear polarized emissions \cite{young2009bell}. To perform the fine structure splitting measurement accurately, the cavity mode was tuned away from the QD transitions to make sure that the cavity does not affect the polarization of the QD emission. Fig.~\ref{f3}(c) shows the fitted energy difference between two peaks as a function of wave plate angle. The solid red line shows the fitted results with sine function. The energy difference oscillates with a period of $\pi$ with an amplitude of 37 $\mu$eV, which is typical for the fine structure splitting energy of InAs QDs \cite{seguin2005size,PhysRevB.69.161301,mar2010electrical,mar2016electrical}. Therefore, we can conclude that peak2 and peak3 originate from XX and X transitions of a single QD respectively.

The contour plot of the temperature dependent PL spectra  is shown in Fig.~\ref{f4}(a). The single-photon Rabi splitting energies of XX-C (Cavity) and X-C polaritons with values over 240 $\mu$eV indicated large $g_{XX}$ and $g_{X}$, leading to reversible energy exchange between cavity and transition even with a large detuning. Due to the proximity between the single-photon Rabi splitting energies and the binding energy, the cavity could simultaneously couple to both XX and X transitions. The dynamics of this model was simulated by solving master equation using Quantum Optics Toolbox \cite{tan2002quantum}. The cavity mode was set to be V-polarized so only coupled to V-polarized transitions (thus $X_H$ was not considered in coupling). The cavity mode was fixed at 1166.3 nm for simplicity and decay rate was 90 $\mu$eV (Q=14000) after deconvolution with our spectrometer's linewidth. XX and X transitions can be quadratically tuned as a function of temperature with a linewidth of 10 $\mu$eV, which was extracted from the experimental data. Due to similarity of two single-photon Rabi splitting energies in our experimental observation, we set $g_X$=$g_{XX}$=$g$ in the calculations. Temperature-dependent PL spectra were simulated with different coupling strengths. In our system $g$ was obtained with value of around 130 $\mu$eV, according to the Rabi splitting energy from the experimental results. The calculation results are shown in Fig.~\ref{f4}(b) on logarithmic color scale, from which two single-photon anti-crossings are observed and additional nonlinearity effects can be resolved in the region between them. Compared with our experimental data in Fig.~\ref{f4}(a), it can be seen that the theoretical calculation result corresponds well with experimental data. Some differences between Fig.~\ref{f4}(a) and Fig.~\ref{f4}(b) are due to that the cavity mode shifts as temperature increases in experiments, while it is kept at same position in the calculation.

The single-photon coupling strength $g$ exceeds the threshold value of $82~\mu$eV in our calculation in Fig.~\ref{f1}(e) (same binding energy and same cavity Q), indicating two-photon Rabi oscillation between $|XX,0\rangle$ and $|G,2\rangle$. The two-photon Rabi splitting could be clearly resolved in calculated PL mapping (Fig.~\ref{f4}(b)). Green (Purple) dashed line schematizes XX-C (X-C) polaritons and XX transition contains two splittings. Splitting at 16 K comes from two transitions XX-P$_{1a}$ and XX-P$_{1b}$, due to the single-photon splitting of $|X,0\rangle$ and $|G,1\rangle$ \cite{winger2008quantum}. While the splitting at 14 K comes from  $P_{2a}$-$P_{1b}$ and $P_{2b}$-$P_{1b}$, due to two-photon Rabi oscillation between$|XX,0\rangle$ and $|G,2\rangle$. Meanwhile the splitting in red lines ($P_{2a}$-$P_{1a}$ and $P_{2b}$-$P_{1a}$) results from the two-photon Rabi oscillation as well. During the two-photon strong coupling, there is a large two-photon emission enhancement region, along with a large suppression region in XX and X transition. Specific calculations and theoretical analysis with different coupling strengths from small to large are shown in Supplemental Materials. These nonlinearity features come from the built-in correlation between XX and X transition of one single QD, which could hardly be observed for a coupled system with one cavity and two different QDs \cite{kim2011strong}.

In our experimental data we could not distinguish every peak shown in theory, which is limited by the linewidth of our spectrometer. However, two-photon Rabi splitting could be proved from temperature-dependent linewidth of peak2 (Fig.~\ref{f4}(c)). The linewidth increases up to a maximum value of $\sim 0.4~nm$ at 11 K and 16 K, because at that temperatures peak2 is combined of several peaks. In the depression region between two maximum values, the minimum linewidth is $\sim 0.3~nm$, much wider than bare cavity mode ($\sim 0. 18~nm$ without deconvolution). The experimental linewidth variation is in good agreement with calculation results (Fig.~\ref{f4}(c) and inset in Fig.~\ref{f4}(b)), considering the broadening of 0.1 nm from spectrometer. In contrast, when $g$ is small and no two-photon Rabi splitting occurs, the minimum linewidth should be almost the same as cavity mode (Fig.~\ref{f4}(c) and Supplemental Materials). Additionally, two suppression regions (insets in Fig.~\ref{f4}(a)) are clearly observed, shown as well in the intensity variation diagram fitted from PL spectra (Fig.~\ref{f4}(d)). The two PL suppression regions (green arrows in Fig.~\ref{f4}(d)) in single-photon emission (peak1 and peak3) are also in good agreement with our theoretical analysis above (two splittings labeled by white arrows in Fig.~\ref{f4}(b)). Suppression at 16 K is due to the single-photon Rabi splitting of XX-$X_V$ transition, with only emission from uncoupled XX-$X_H$ transition left (not shown in simulation result). Large suppression region around 13 K results from strong coupling along with emission enhancement of two-photon process, correspondingly. In contrast, when $g$ is small, this suppression region should reduce to a point due to weak coupling in two-photon process \cite{ota2011spontaneous}. On account of simulation and analysis of the experimental data, we can confirm the two-photon Rabi splitting in our coupled biexciton-cavity system.

In conclusion, we demonstrated two-photon Rabi splitting in a strongly coupled system consisting of an L3 PC cavity and a single embedded QD with multiple exciton states. Both XX and X transitions of the single QD were strongly coupled to one cavity mode with large coupling strengths of 130 $\mu$eV. Such a large coupling strength close to half of the binding energy 350 $\mu$eV enabled the cavity to simultaneously couple to two single-photon transitions, leading to two-photon Rabi splitting between biexciton and cavity mode as predicted by theoretical analysis and simulation. Our work promotes the strong coupling regime in cavity-dot system from single-photon process to multi-photon process, providing an approach for multi-qubit operation. Additionally, our cavity-dot system can be easily integrated with PC waveguides with a wavelength approaching the telecommunication regime, which has great potential for quantum photonic network.

\begin{acknowledgments}
This work was supported by the National Basic Research Program of China under Grant No. 2014CB921003; the National Natural Science Foundation of China under Grant No. 11721404, 51761145104 and 61675228; the Strategic Priority Research Program of the Chinese Academy of Sciences under Grant No. XDB07030200 and XDPB0803, and CAS Interdisciplinary Innovation Team. Authors would like to thank Gas Sensing Solutions Ltd for using the MBE equipment.
\end{acknowledgments}

\end{document}